# Characterizing Exoplanetary Atmospheres

Jonathan J. Fortney[1]

[1]*Department of Astronomy and Astrophysics, University of California, Santa Cruz, Santa Cruz, CA, 95064, USA. jfortney@ucsc.edu

**Abstract**

I review the major open science questions in exoplanet atmospheres. These are mainly focused in the areas of understanding atmospheric physics, the atmosphere as a window into other realms of planetary physics, and the atmosphere is a window into understanding planet formation. For gas giant planets, high quality spectra have been delivered from *JWST* and from the ground, enabling the determination of atmospheric abundances. For the very common sub-Neptune planets, we are just beginning to obtain and interpret *JWST* spectra. For the terrestrial planets, which can be studied only around M stars, the field aims to determine if these planets even have long-lived atmospheres.

## Introduction

The late 1990s saw a number of modeling efforts to begin to understand exoplanetary atmospheres [1–3], and in 2002 Charbonneau et al. [4] used *Hubble* to detect sodium atoms in the atmosphere of the transiting hot Jupiter exoplanet, HD 209458b, a 1500 K planet on a 3.5 day orbit. The field has dramatically expanded since that time, in terms of the range of modeling efforts, the number, quality, and wavelength range of spectra achieved, and the number of people working in the field. All of these areas continue to grow rapidly today, in many ways powered by *JWST*. Exoplanetary atmospheres has a truly tremendous amount of undiscovered country, with nearly all basic questions still to be uncovered. It can be a bit overwhelming. In order to set the scene, I think it is important to categorize the kinds of questions we are trying to answer, today, and how these kinds of questions differ depending on the type of planet.

**How do planets work?** What is the physics and chemistry operating within planetary atmospheres? How do atmospheric spectra tell us about the physics of the atmospheres as well as the physics of the time-evolving planetary interior far below the visible atmosphere? The abundances of some atoms and molecules may be representative of the local temperature and pressure conditions within the atmosphere. However, some molecules may instead be signposts of conditions far below the visible atmosphere, mixed vigorously upwards on timescales faster than chemical reaction timescales [5], or altered via incident stellar UV flux that drives photochemistry. Still other molecules, via their presence or absence, may be signposts of physical or chemical processes at play between a deep atmosphere and a solid or liquid surface. In much the same way that chemical abundances in Sunlike and evolved stars are used to constrain current and past interior processes, the same is true for planets.

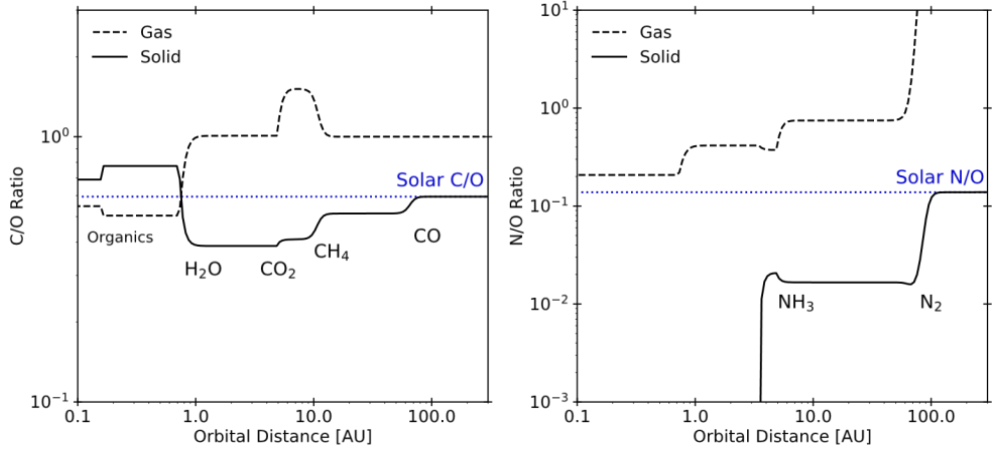

**Fig. 1** The C/O ratio (left) and N/O ratio (right) for nominal 1D protoplanetary disks. The solar ratios are in dotted blue. Condensation at a "snow line" of various volatile molecules (e.g., $H_2O$, $CO_2$) alters the C/N/O ratios of both the condensed solids and the leftover disk gas. Generally larger effects are seen for N/O than C/O. A giant planet will accrete both gas and solids to make up its metal-rich H/He dominated envelope and atmosphere. Figure courtesy Kazumasa Ohno, after [6], expanded from [7].

**How do planets form?** Do atmospheric abundances hold clues to aspects of a planet's formation? Since a planet-forming disk, in bulk, has the same composition as that of the parent star, any deviations in planetary composition (more precisely giant planets, which accrete nebular H/He) may yield important insights into protoplanetary disks. Giant planets accrete both gas and solids [8], and the composition of both gas and solids change with orbital separation within the disk, in particular at condensation points, or "snow lines" where compounds change from vapor to solids. The relative abundance of important molecules like C, N, O, or other elements may inform aspects of the formation location and relative accretion of gas and solids, as depicted in a simple disk model in Figure 1. For planets less than ∼5 $M_\oplus$, particularly those on close-in orbits, an open question is whether some of these planets have atmospheres at all, given the high bolometric luminosity and XUV fluxes for their parent stars at young ages.

## Methods of Observations

Photometry and spectroscopy of exoplanets, for transiting planets, and for directly imaged planets, both grew up quickly starting in the mid-2000s. For transits, *Spitzer* (in particular *Warm Spitzer*) and *Hubble* STIS and WFC3 provided photometry and limited-wavelength spectroscopy, respectively, for a few dozen planets. For imaged planets, the Gemini Planet Imager, SPHERE, VLT/GRAVITY, and Keck (among others) provided mostly JHK spectroscopy for around two dozen planets.

## Transit Spectroscopy

Transiting planets are observed at a number of possible orbital phases. During the transit stellar photons pass through the day-night "terminator" transition region of the planetary atmosphere, imprinting planetary atmosphere absorption features on the spectrum. The size of these features are small – the area of the annulus of planetary atmosphere compared to the stellar cross-sectional area. For a given transit depth $\delta$ the extra transit depth $\Delta\delta$ due to atmospheric absorption is:

$$\delta = \frac{R_p^2}{R_\star^2}, \quad \Delta\delta \approx 2 N_H(\lambda)\delta \left(\frac{H}{R_p}\right) \propto \left(\frac{H R_p}{R_\star^2}\right) \tag{1}$$

where $R_p$ and $R_\star$ are the planetary and stellar and stellar radii, respectively, $N_H$ is the number of atmospheric scale heights probed (typically a few, and depends sensitively on the wavelength dependent opacity), and $H$ is the atmosphere's scale height. A detailed study of the time-evolving ingress and egress lightcurves, at the start and end of the transit, may allow for the determination of differences in chemistry and temperatures of leading and trailing hemispheres of the planet. These properties constrain the hydrodynamics of atmospheres.

Transiting planets typically (but not always, if their orbits are eccentric) have a secondary eclipse, or occultation, when the planet passes directly behind the parent star. The disappearance of the planet's thermal emission, or potentially reflection, can be measured as well. For hot planets this signal can be larger than the atmosphere transmission signal, while also giving details on atmospheric thermal structure, but for cool planets the signal is challenging. For blackbody-temperature planetary ($T_p$) and stellar ($T_\star$) thermal emission, the size of the signal $\delta_{occ}(\lambda)$ is:

$$\delta_{\rm occ}(\lambda) = \delta \, \frac{B_\lambda(T_p)}{B_\lambda(T_\star)} \longrightarrow \delta \, \frac{T_p}{T_\star} \tag{2}$$

where the portion after the arrow indicates the further simplification of assuming the Rayleigh-Jeans limit. A detailed study of the time-evolving ingress and egress lightcurves, as the planet is gradually occulted, may allow for day-side thermal emission "eclipse maps," which provide a snapshot of the inhomogeneity of the planetary day side, including temperature or cloud-cover differences.

Transiting planet phase curves are also an important observational tool [9]. By measuring the thermal emission over an entire orbital period for the planet, the planet's day side emission, night side emission, and any hot and cool spot phase shifts in longitude can be measured. Via spectroscopy, these shifts ban be measured as a function of wavelength, and hence, depth in the atmosphere, yielding a comprehensive 3D view of circulation in visible parts of the atmosphere.

## Imaging Spectroscopy

For directly imaged planets, the difficulty in obtaining data is the twin problems of the necessary contrast, and the inner working angle of your telescope. The current state of the art is a contrast ratio of around $10^{-6}$ in flux. That being said, spectroscopy of these planets spans a broad range of spectral resolution (from $R \sim 100$ to $\geq 10{,}000$ and signal-to-noise ratio. Nearly all of these objects are young (tens of Myr or younger) and massive (more than a Jupiter mass) as older or smaller planets are just too faint. *JWST* may provide spectra for smaller and fainter planets on wide separations from M dwarfs or white dwarfs. A current area of interest is time series spectroscopy over an entire planetary rotation period ($\sim$ 5-15 hours). The time-variable emission can inform our understanding of cloud coverage (static or also itself time-varying) as a function of planetary longitude. This time-variable emission has been measured for many isolated brown dwarfs, in the absence of a bright parent star [10].

# Main Physical Ideas

## Planet Types

Gas giant planets like Jupiter (318 $M_\oplus$) have most of their mass in the massive H/He dominated envelope that they accrete from the protoplanetary nebula. The visible atmosphere is the top of this envelope. At lower masses, one transitions to "sub-Saturns" or "super-Neptunes," ($\sim 20 - 75$ $M_\oplus$?) where the mass of H/He and the mass of total metals (in the central core and in the H/He envelope) are comparable. For Neptune and Uranus in particular ($\sim 15$ $M_\oplus$), their H/He mass is only 10-20% of the total mass. For still smaller planets, the "sub-Neptunes" or "mini-Neptunes," the H/He envelope is $\sim$ 0.1–5% of the planetary mass and these planets are typically 2-3 $R_\oplus$ and 5-10 $M_\oplus$. The *Kepler* mission showed that, at least inside of 100 day orbits, sub-Neptunes dominate over Neptune-mass and Jupiter-mass planets. A prediction from planet formation theory is that the metallicity of H/He-dominated atmospheres will be strongly anticorrelated with mass, which is seen in the solar system's 4 giant planets [11], where Jupiter has the lowest metallicity, of around 3× solar.

At some point in mass/temperature space ($\sim 5$ $M_\oplus$???), planets transition from having accreted H-dominated atmospheres to those dominated by $H_2O$ (perhaps with liquid water beneath: "water worlds"), $CO_2$ (like Venus and Mars), or $N_2$, like Earth, or perhaps other variations. These higher mean molecular weight "secondary atmospheres" are outgassed from the interior, rather than the H/He-dominated "primary atmospheres" accreted from the protoplanetary nebula. In the solar system there are clear distinctions between primary and secondary atmospheres, but that need not be the case, in particular for the sub-Neptunes. Similar-mass planets with secondary atmospheres, or perhaps no atmospheres at all, are usually termed "super-Earths."

## Atmospheric Energy Balance

There are three important characteristic temperatures that come up often when discussing planetary atmospheres: effective temperature ($T_{\text{eff}}$), which characterizes the total thermal output of the planet, equilibrium temperature ($T_{\text{eq}}$), which characterizes the thermal output due only to absorbed and re-radiated stellar energy, and interior temperature ($T_{int}$), which characterizes the thermal energy only from the cooling of the planet's interior. These temperatures are related by:

$$T_{eff}^4 = T_{eq}^4 + T_{int}^4 \tag{3}$$

For a small old planet like Earth, $T_{int}$ is negligible, and $T_{eff} \approx T_{eq}$. For a close-in hot Jupiter exoplanet, the incident stellar energy is extreme, driving $T_{eff} \approx T_{eq}$ as well, even if $T_{int}$ is several hundred K. However, for a young gas giant planet on a wide separation orbit, $T_{int}$ is large (typically 1000+ K for known imaged planets), so that $T_{eff} \approx T_{int}$.

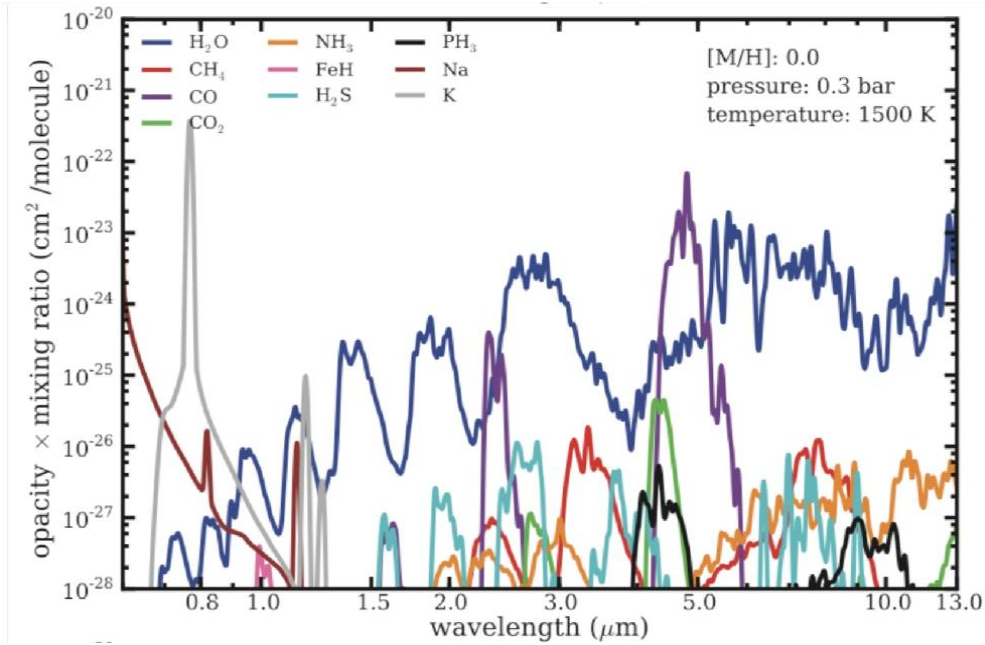

**Fig. 2** The expected dominant atmospheric opacity sources for hot giant atmospheres. The plot is at 1500 K and 0.3 bar, with spectral resolution $R \sim 100$. It shows the absorption cross-section multiplied by the expected volume mixing ratio of each gas species at solar metallicity. *JWST* wavelengths capture the major absorption features of K, $H_2O$, CO, $CH_4$, $CO_2$, and $NH_3$. High metallicities and temperatures tend to favor CO and $CO_2$, and cooler temperatures favor $CH_4$ and $NH_3$, compared to this particular plot. Plot

## Main Atmospheric Opacity Sources

The atmospheres of planets are typically dominated by the opacity of molecules, some atoms, and clouds. In giant planets, $H_2O$, CO, $CO_2$, $CH_4$, $NH_3$, $SO_2$, $H_2S$, Na, and K, are seen across a range of hot and warm planets from 700–2000 K with *JWST*, as shown in Figure 2. The very hottest planets have TiO, VO, and a host of atomic metals, like M dwarfs. All of these species have strongly wavelength dependent opacities, so that there is no "continuum," and consequently a wide range of atmospheric pressures are probed both in emission and transmission spectroscopy. The infrared spectroscopic capabilities of *JWST* are well-matched to the main absorption features of a host of important molecules.

For exoplanet atmospheres, clouds are like the "magnetic fields" of the rest of astrophysics – extremely important, difficult to model, and ignored only at your own risk. For the hot Jupiters at $T_{eff} \sim$ 1500–2000 K, the dominant cloud species are expected to be silicate rock dust and iron droplets. We have ample evidence for the silicate clouds in brown dwarfs of similar temperatures. At cooler temperatures, $T_{eff} \sim$ 1000 K, a host of other elements begin to condense, and $Na_2S$ and KCl may form potentially optically thick clouds, especially as high metallicity. Eventually water clouds will form at $T_{eff} \sim$ 400 K.

Another type of condensates, photochemical hazes, are certainly seen in most solar system atmospheres, and dominate the atmospheric opacity of Saturn's moon Titan. Our understanding of these aerosols in still in its infancy, as we cannot use the better studied (but isolated) brown dwarfs as any sort of guide for giant planet hazes. Our solar system's terrestrial planets likely cover only a modest phase space of possible hazes compared to exoplanets. It has been suggested that range of possible haze compositions in exoplanetary atmospheres is far larger than in the solar system, and there is a growing body of work demonstrating this in the laboratory.

## The Major Science Questions

- How do giant planet atmospheres differ in composition compared to their parent stars? The solar system sample is limited by a tiny sample size and limited number of elements. What is the degree of metal-enrichment and non-stellar abundance ratios? (While many/most appear metal-rich to varying degrees, at least one giant planet atmosphere actually appears *metal-poor*! [12])
- Do terrestrial planets around M dwarfs have atmospheres? The prolonged high luminosity phase of pre-main sequence M dwarfs may drive the atmospheres off of planets that today would be able to sustain atmospheres. There is likely to be a major push in this area in *JWST* cycles 3-5, to find the "Cosmic Shoreline," as suggested in Figure 3. Whether these planets have atmospheres, or not, has major ramifications for ELT plans for exoplanet direct imaging, which seeks to study reflection and thermal emission of M-dwarf terrestrial planets.
- What is the nature (natures?) of sub-Neptune exoplanets? Their atmospheres will tell us if this population is dominated by those with thin H/He envelopes or if some are water-worlds. Chemical abundances may hold clues to deep water or magma layers, below H/He- or steam-dominated atmospheres due to atmosphere/interior chemistry [13] affecting visible atmospheric abundances. What do combination primary/secondary atmospheres (unseen in the solar system) look like?
- For what kind of planets, and over what temperatures and compositions, do aerosols have a major impact on atmospheric opacity? This is important for a range of physics and chemistry issues, but also for the proposed Habitable World Observatory (HWO), where the reflected light signal is quite sensitive to clouds for terrestrial and giant planets.

- Can we understand the mechanisms that drive the atmospheric variability that is well-known in brown dwarfs, strongly suspected to occur in wide-separation imaged planets, and may also be detectable for transiting planets with strong day/night contrasts? The atmospheric dynamics of these objects span a wide range of physical regimes in rotation rate, surface gravity, and thermal forcing. [14]
- How well can we extract information about inherently 3D atmospheric properties of exoplanets from their 1D disc- and terminator averaged observations? What are the limits on accuracy and precision for exoplanet spectroscopy, in emission, transmission and reflection, for planetary photospheres that can be strongly inhomogeneous?

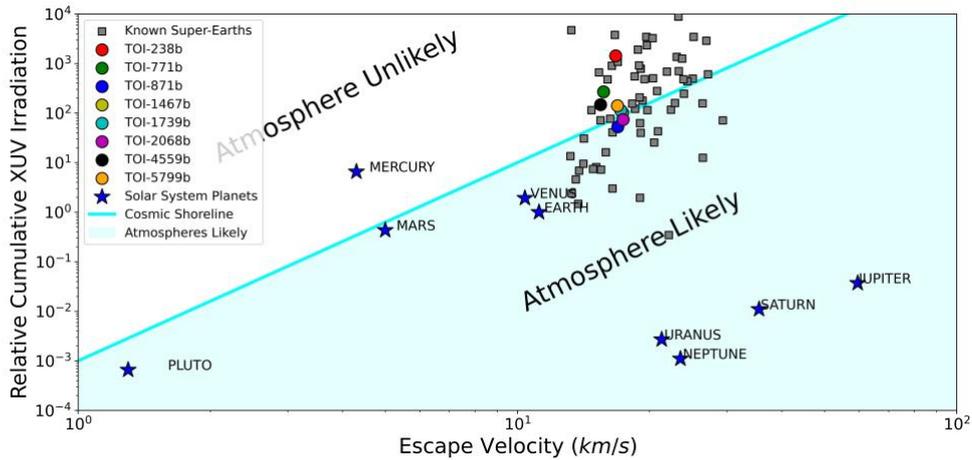

**Fig. 3** Cumulative XUV irradiation vs. planetary escape velocity for a range of rocky exoplanets (grey squares and colored circles) and solar system planets (blue stars). The cumulative lifetime XUV is thought to be the driving force behind planetary evaporation. Atmospheres can be found where gravity is high and solar irradiation is low. The light blue colored line represents the proposed "cosmic shoreline" which follows a power law, $I_{XUV} \propto v_{esc}^4$, following [15]. Rocky planets around M dwarfs can be studied with *JWST* to assess the validity of this shoreline. Plot adapted from [16].

## Conclusions

The exoplanet atmospheres field is rapidly changing with the current flood of new *JWST* observations. For giant planets, a comprehensive theory of their atmospheres, that crosses from transiting planets, to imaged planets, to the solar system, and brown dwarfs, across a range of $T_{eff}$, surface gravity, and formation conditions, seems like an attainable goal over the coming years. For the sub-Neptunes, which dominate the *Kepler* and *TESS* exoplanet populations, we await spectra across a range of planet masses, orbital separations, and incident flux levels. The complexity of these atmospheres is sure to be significant, given the role of the atmosphere/interior composition connection in altering molecular abundances. For rocky planets around M dwarfs, we are starting to build up a sample of planets that do or do not have atmospheres, via transiting planet transmission and emission spectroscopy. Understanding which of these planets have atmospheres, and the nature of these atmospheres, will be a significant undertaking over the next several years.

**Acknowledgments.** I wish to thank Dr. Zhoujian (Z.J.) Zhang and Dr. Brianna Lacy for comments on the draft. Additional great thanks to Dr. Ewine van Dishoeck for the invitation to attend the meeting.